\documentclass{emulateapj}
\usepackage{apjfonts}

\begin{document}

\shortauthors{Berdyugina et al.}
\shorttitle{First direct detection of scattered light from an exoplanet}

\title{First detection of polarized scattered light from an exoplanetary atmosphere}

\author{S. V. Berdyugina\altaffilmark{1,2}, 
A. V. Berdyugin\altaffilmark{2}, 
D. M. Fluri\altaffilmark{1}, 
V. Piirola\altaffilmark{2,3} }

\altaffiltext{1}{Institute of Astronomy, ETH Zurich, CH-8092 Zurich, Switzerland;
sveta@astro.phys.ethz.ch, fluri@astro.phys.ethz.ch }

\altaffiltext{2}{Tuorla Observatory, University of Turku, V\"ais\"al\"antie 20, FIN-21500, Piikki\"o, Finland; andber@utu.fi, piirola@utu.fi }

\altaffiltext{3}{Vatican Observatory, V-00120 Citta del Vaticano}

\begin{abstract}
We report the first direct detection of an exoplanet in the visible polarized light. The transiting planet HD189733b is one of the very hot Jupiters with shortest periods and, thus, smallest orbits, which makes them ideal candidates for polarimetric detections. We obtained polarimetric measurements of HD189733 in the $B$ band well distributed over the orbital period and detected two polarization maxima near planetary elongations with a peak amplitude of $\sim$2$\cdot$10$^{-4}$. Assuming Rayleigh scattering, we estimated the effective size of the scattering atmosphere (Lambert sphere) to be 1.5$\pm$0.2\,$R_{\rm J}$, which is 30\%\ larger than the radius of the opaque body previously inferred from transits. If the scattering matter fills the planetary Roche lobe, the lower limit of the geometrical albedo can be estimated as 0.14. The phase dependence of polarization indicates that the planetary orbit is oriented almost in a north-south direction with a longitude of ascending node $\Omega$=(16\degr\ or 196\degr)$\pm$8\degr. We obtain independent estimates of the orbit inclination $i$=98\degr$\pm$8\degr and eccentricity $e$=0.0 (with an uncertainty of 0.05) which are in excellent agreement with values determined previously from transits and radial velocities. Our findings clearly demonstrate the power of polarimetry and open a new dimension in exploring exoplanetary atmospheres even for systems without transits.
\end{abstract}


\section{Introduction}

More than 200 extrasolar planets have been discovered by the indirect methods of Doppler spectroscopy, photometric transits, microlensing, and pulsar timing \citep[e.g.][]{marcy05,udry_santos06}. However, the direct detection of exoplanets, enabling a study of their physical properties, remains a challenge. Thus far atmospheres have been detected in transiting planets \citep[e.g.][]{char02,dem06} and only in one non-transiting planet $\upsilon$~And observed in the far infrared with {\it Spitzer} \citep{har06}.
Polarimetry is praised as a powerful technique \citep{hou06,kel06,sch06} for detecting directly starlight that is scattered in a planetary atmosphere and, thus, possesses information on its geometry, chemistry, and thermodynamics. However, due to the anticipated very low polarization degree of $\sim$10$^{-5}$, 
a practical use of this method was generally attributed to the future. Here we report the first polarimetric detection of an exoplanet at a level exceeding theoretical expectations by an order of magnitude, primarily due to the choice of the recently discovered hot Jupiter HD189733b \citep{bou05} and the fact that the scattering planetary atmosphere is found to be somewhat larger than assumed before.

The light scattered in the planetary atmosphere is linearly polarized perpendicular to the scattering plane. It is best characterized by the Stokes parameters $q$ and $u$, normalized to the total flux. In general, when the planet revolves around the parent star, the scattering angle changes and the Stokes parameters vary. If the orbit is close to circular, two peaks per orbital period can be observed. The observed polarization variability should thus, in principle, exhibit the orbital period of the planet and reveal the inclination, eccentricity, and orientation of the orbit, and, if detected with high enough polarimetric accuracy, also the nature of scattering particles in the planetary atmosphere.

Thus, it appears feasible to detect the scattered light with polarimetric 
methods, even if an exoplanet is not spatially resolved. Hot Jupiters orbiting their host stars on close orbits are therefore the best candidates for the first detection. Due to their proximity to the star, such planets apparently develop 
extended hydrogen halos \citep{vm03}, which may effectively scatter the 
stellar light, especially in blue wavelengths. 

In this Letter we report the first direct detection of an extrasolar planet in the visible light. We present results of our polarimetric monitoring of the known exoplanet HD189733b in the $B$ band. A large number of measurements allows us to reduce the statistical error and for the first time detect directly the light scattered in the atmosphere of an extrasolar planet. We interpret the data employing a model based on the Lambert sphere approximation and Rayleigh scattering \citep{flu07} and deduce the radius of the sphere and the orbit orientation on the sky plane. 

\begin{figure}
\resizebox{8.5cm}{!}{\includegraphics{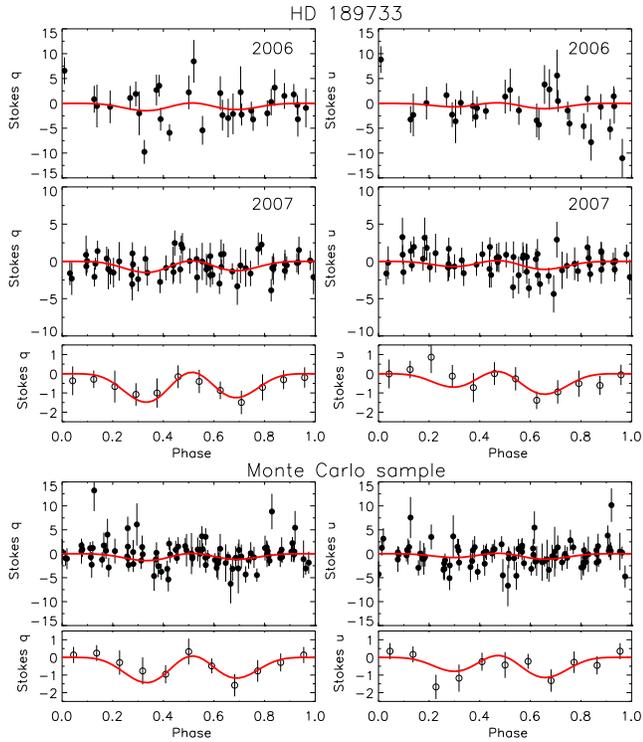}}
\caption{Polarimetric data (Stokes $q$ and $u$ with $\pm1\sigma$ error bars
in the scale of 10$^{-4}$) for HD189733 and a Monte Carlo simulated sample. 
The original data are shown with {\it filled circles}: for HD189733 in the top row panels for the year 2006 and in the middle row panels for 2007, and for the Monte Carlo simulated data combined for both years in single panels. The data rebinned for equal phase intervals are shown with {\it open circles} in separate panels. For HD189733, the constant shifts  in Stokes parameters $\Delta q$ and $\Delta u$ were subtracted from the data according to Table~\ref{tab:par}. The best-fit solutions deduced from the unbinned data are shown with {\it solid curves}. [{\it See the electronic edition of the Journal for a color version of this figure.}] 
}
\label{fig:pol}
\end{figure}

\section{Observations}\label{sec:obs}

To increase the probability for detecting polarized scattered light from an exoplanet, we selected the nearby system HD189733 with the planet at a short distance from the star (0.03\,AU) and, thus, with a short orbital period (2.2\,days). It was recently discovered in the ELODIE search for transiting planets \citep{bou05} and immediately became one of the favorite targets. 

Our observations were carried out in 2006--2007 with the double image CCD polarimeter DIPol \citep{piirola05} installed on the remotely controlled 60 cm KVA telescope on La Palma, Spain.  It is equipped with a rotating superachromatic half-wave plate as the retarder and a calcite plate as the analyzer. 
To measure linear polarization, the retarder was rotated at 22.5\degr\ intervals. Each pair of the observed Stokes $q$ and $u$ was calculated from four exposures at different orientations of the retarder. To avoid systematic errors, the observations were carried out in cycles of 16 exposures, corresponding  to a full rotation of the retarder.
In the 2006 season we made 10--15\,s exposures at $2\times16$ positions, yielding eight single observations of $q$ and $u$ per night. These were then averaged to calculate the nightly mean value and its standard error (1$\sigma$).
Typical errors of the 2006 measurements were 0.02--0.03\%.
In 2007, in order to reduce the measurement errors, we increased the integration time for invidual exposures up to 20--30\,s and made measurements at $4\times16$ positions, thus increasing the total integration time by a factor of 4. This reduced the errors by a factor of 2, down to 0.01--0.015\%, which indicates that the accuracy was limited by the photon noise and did not suffer from systematic effects. Overall we obtained 93 nightly measurements for each Stokes parameter. This allowed us to reduce the statistical error down to 0.006\%\ on average in the binned data (see Fig.~\ref{fig:pol}) and clearly reveal polarization peaks of $\sim$0.02\%\ near elongations.

\begin{deluxetable}{lcc}
\tabletypesize{\scriptsize}
\tablewidth{8.5cm}
\tablecaption{Parameters of the HD189733 System.
\label{tab:par}}
\tablehead{
\colhead{Parameter} & \colhead{Known value} & \colhead{Best fit value} }
\startdata
P, days             & 2.218581$^{(1)}$    & ...\\
T$_0$, JD2,400,000+ & 53931.12048$^{(1)}$ & ...\\
$R_*$/$R_\sun$      & 0.76                & ...\\
$a$, AU             & 0.0312              & ...\\
$e$                 & 0.0                 & 0.0$^{(2)}$ \\
i\degr\             & 85.68$^{(1)}$       & 98$\pm$8      \\
$\Omega\degr$       &  ...                & 16(196)$\pm$8     \\
$\Delta q$/10$^{-4}$&  ...                & $-$2.0$\pm$0.3 \\
$\Delta u$/10$^{-4}$&  ...                & $-$0.7$\pm$0.3 \\
$R_{\rm L}$/$R_{\rm J}$ & 1.15$^{(1)}$    & 1.5$\pm$0.2    \\
$M/M_{\rm J}$       & 1.15                & ...\\
$R_{\rm RL}$/$R_{\rm J}$ & 3.3            & ...\\
$p_{\rm RL}$        & ...                 & 0.14 \\
\enddata
\tablecomments{
$^{(1)}$ \citet{pont07}. Other fixed parameters are from \citet{but06}.
$^{(2)}$ The uncertainty of $e$ is 0.05.
}
\end{deluxetable}

For calibration of the polarization angle zero point we observed the highly polarized standard stars HD204827 and HD161056. To estimate the value of the instrumental polarization, a number of zero polarized nearby ($<$25\,pc) stars from the list by \citet{piirola77} were also observed. In fact, the instrumental polarization at the KVA telescope has been monitored since 2004 within other projects as well \citep[e.g.,][]{piirola05}. These measurements demonstrated that in the $B$-passband it was well below 0.02\%\ and invariable. 

\section{Modeling}\label{sec:mod}

To analyze the observed polarimetric signal, we employ a simple model based on the Lambert sphere approximation, i.e., a perfectly reflecting surface with the geometrical albedo $p=$2/3, and Rayleigh scattering \citep{flu07}. Modeling the observed variations in Stokes $q$ and $u$ allows us to reconstruct the orientation of the planetary orbit
in space and estimate the effective size of the scattering atmosphere (Lambert sphere). In the model, fixed parameters are the orbital period $P$, transit or periastron epoch $T_0$, semi-major axis $a$, and the radius of the star $R_*$, which is considered to be a limb-darkened sphere. The values used are provided in Table~\ref{tab:par}. The limb-darkening was assumed according to \citet{claret00}, but its details were found to be insignificant within the measurement errors.
Free parameters are the eccentricity $e$, orbit inclination $i$, longitude of the ascending node $\Omega$, radius of the Lambert sphere $R_{\rm L}$, and constant shifts in Stokes parameters $\Delta q$ and $\Delta u$, which can be present in the data due to interstellar or circumstellar polarization. In the case of transiting planets, the orbit inclination can also be determined from photometric data, which is a valuable test for our model. Otherwise, polarimetry provides a unique opportunity to evaluate both $i$ and $\Omega$. Moreover, it is possible to distinguish between inclinations smaller and larger than 90\degr, which is not possible from transit data. 

In general, it is $R_{\rm L}$ that scales the amplitude of polarization variations. The inclination scales the relative amplitudes in Stokes $q$ and $u$. For example, at $i=0$\degr\ $q$ and $u$ have the same amplitude. If $i\ne0$\degr, the relative amplitude is also influenced by $\Omega$, e.g., at $i=90$\degr\ variations appear only in Stokes $q$ if $\Omega=0$\degr, 90\degr, 180\degr, or 270\degr, and only in Stokes $u$ if $\Omega=45$\degr, 135\degr, 225\degr, or 315\degr. More examples can be found in \citet{flu07}. Observed polarization can be both positive and negative, since its direction is always perpendicular to the line joining the planet and the star as projected on the sky plane. Our definition is in accordance with the common assumption that positive $q$ is in the north-south direction, while the negative one in the east-west direction. Positive and negative $u$ are at an angle of 45\degr\ counterclockwise from the positive and negative $q$, respectively (see Fig.~\ref{fig:orb}). The inclination is defined in such a way that the planet revolves counterclockwise as projected on the sky for 0\degr$\le i<$90\degr\ and clockwise for 90\degr$< i\le$180\degr. Further, $\Omega$ varies from 0\degr\ to 360\degr\ starting from the north and increases via east, south, and west.

In many cases, two maxima per period near the elongations are expected. For an orbit with moderate eccentricity the maxima shift closer to the periastron. Our polarimetric data reveal a clear periodic modulation with an amplitude of $\sim$2$\cdot$10$^{-4}$ in Stokes $q$ and about half of that in Stokes $u$, with the polarization being negative, indicating the $\Omega$ value to be near either 0\degr\ or 180\degr\ at high inclination, which is confirmed by the model.

\begin{figure}
\resizebox{8.5cm}{!}{\includegraphics{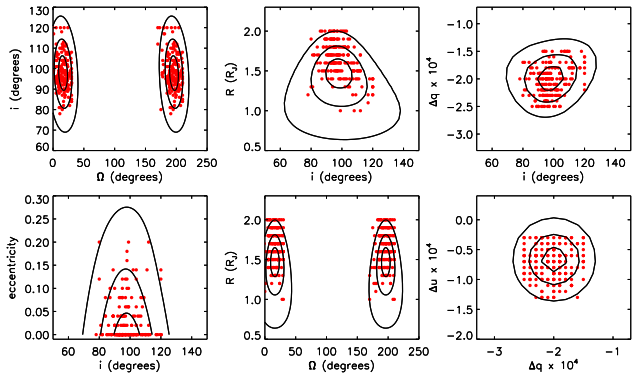}}
\caption{The $\chi^2$ contours of the best-fit solution for HD189733b ({\it solid lines}). The contours are shown for the confidence levels 68.3\%, 90.0\%, and 99.0\%. Due to the intrinsic ambiguity in $\Omega$ two equal minima separated by 180\degr\ are seen. The Monte Carlo sample solutions are shown with dots. [{\it See the electronic edition of the Journal for a color version of this figure.}]}
\label{fig:chi}
\end{figure}

The model parameters are estimated using a $\chi^2$ minimization procedure applied to the bulk of the original measurements for both years 2006 and 2007 simultaneously. Tests with simulated data showed that for fixed values of $P$ and $T_0$ and reasonably large number of measurements ($>$50) the procedure is able to find a unique solution even for low signal-to-noise ratios \citep{flu07}. The only ambiguity is in the $\Omega$ value of 180\degr, which is due to the intrinsic property of Stokes $q$ and $u$ to remain unchanged under the rotation by 180\degr. The best-fit values of the parameters found at the normalized $\chi^2$=1.04 are listed in Table~\ref{tab:par}. We find an excellent agreement with known values of eccentricity and inclination. Interestingly, the inclination of 98\degr\ revealed by polarimetry is very close to the complementary value 180\degr--86\degr=94\degr\ inferred with the transit method, which cannot distinguish between inclinations smaller and larger than 90\degr. However, since the Stokes $u$ amplitude is quite small, a better polarimetric accuracy is needed to distinguish between 86\degr\ and 94\degr. The fits to the observations and $\chi^2$ contours of the solution are shown in Figs.~\ref{fig:pol} and~\ref{fig:chi}, respectively. Note that the binned data also follow well the model, thus, indicating that the error distribution is close to Gaussian and the binned data reveal the true variations.

To evaluate the robustness of the solution to the measurement errors as well as to properly evaluate the errors of the model parameters, we employed the Monte Carlo method. Following the recipe by \citet{press94}, we assumed the true orbit to be the same as our best-fit solution and simulated 200 samples of measurements (found to be enough for convergence) at the same orbital phases and with the same errors as the data for HD189733. One sample is shown in Fig.~\ref{fig:pol}. We applied our $\chi^2$ minimization procedure to the simulated samples and obtained best-fit solutions for them. This allows us to obtain the distribution of deviations from the true set of parameters caused by measurement errors. In Fig.~\ref{fig:chi} the best-fit solutions for the simulated samples are plotted together with the $\chi^2$ contours for the HD189733 solution. Impressively enough the Monte Carlo tests concentrate near the $\chi^2$ minimum for all parameters, which proves that (1) the errors of our measurements have a Gaussian distribution, (2) the signal is not spurious, and (3) the solution is robust to the errors. Thus, we are able to evaluate the uncertainties of the HD189733 parameters using the Monte Carlo deviations. These are provided in Table~\ref{tab:par}.

\section{Properties of HD189733b}\label{sec:res}

HD189733b is known as a 'very hot Jupiter'. Its radius in the $B$ band and the orbit inclination were first determined as $R$=1.26$\pm$0.03\,$R_{\rm J}$ and $i$=85.3\degr$\pm$0.1\degr\ \citep{bou05}. Interestingly, the $B$ band transit was observed 20\%\ deeper than the transit in the $R$ band, implying a larger radius in blue wavelengths. Improved parameters 
were most recently obtained by \cite{pont07}: $R$=1.154$\pm$0.017\,$R_{\rm J}$ and $i$=85.68\degr$\pm$0.04\degr. However, the above estimates of the planet radius were obtained by measuring very accurately the planet-to-star diameter ratio and evaluating the stellar radius from the spectral class and colors with some uncertainty. To overcome this uncertainty, \cite{baines07} measured the stellar disk directly using the CHARA Array in the near-infrared $H$ band and yielded for the planet 1.19$\pm$0.08\,$R_{\rm J}$. The direct detection of the thermal emission with {\it Spitzer} using the secondary eclipse indicated a 16\,$\mu$m brightness temperature of 1117$\pm$42\,K \citep{dem06} for the radius estimated by \citet{bou05}. Very recently an infrared continuum spectrum with 0.49\%$\pm$0.02\% of the flux of the parent star \citep{gri07} and perhaps water vapour \citep{tin07}
were detected with {\it Spitzer}.

The radius of the Lambert sphere $R_{\rm L}$=1.5$\pm$0.2\,$R_{\rm J}$ inferred by our modeling is $\sim$30\%\ larger than the radius of the opaque body revealed in the optical, although the lower 1$\sigma$ margin is comparable with the latter and especially with the value by \citet{bou05} for the $B$ band. Also, the Monte Carlo simulations indicate that the inferred value of $R_{\rm L}$ might be biased from the true value by $\sim$0.1--0.2\,$R_{\rm J}$ due to an asymmetric shape of the minimum of the $\chi^2$ distribution (see Fig.~\ref{fig:chi}). The uncertainty in the radius is largely defined by the error in determining the polarization amplitude. In addition, since zero polarization at all orbital phases would correspond to the zero scattering radius (when the albedo value is not zero), measurement errors drag the solution to positive values and cause the asymmetry of the $\chi^2$ minimum.
Nevertheless, the Lambert sphere radius corresponds to the largest possible geometrical albedo of 2/3, which implies that in reality the radius of the scattering atmosphere might be larger than 1.5\,$R_{\rm J}$. For instance, if the geometrical albedo of HD189733b is comparable to those of the giant planets in the Solar system, e.g., 0.52 for Jupiter and 0.47 for Saturn, then the scattering radius of HD189733b would be $R_{\rm L}$=1.7$\pm$0.2\,$R_{\rm J}$. A larger scattering atmosphere supports the idea of extended, evaporating halos around hot Jupiters. If the evaporation does take place, as in the case of, e.g., HD209458b \citep{vm03}, the size of the scattering surface might be comparable with the planetary Roche lobe size, $R_{\rm RL}$. We estimate $R_{\rm RL}$ for HD189733b to be 3.3\,$R_{\rm J}$ \citep{pac71} and the corresponding lower limit of the geometrical albedo  
$p_{\rm RL}=2/3(R_{\rm L}/R_{\rm RL})^2$=0.14.
Another explanation for the detected polarization amplitude could be a possible oblateness of the planet \citep{sen06}.

An excellent agreement between the known and inferred values of the eccentricity and orbit inclination (Table~\ref{tab:par}) indicates the plausibility of the assumed Rayleigh scattering, as its angular dependence strongly influences the shape of the polarization phase curve \citep[e.g.,][]{seager00,stam04}. Therefore, the scattering most probably occurs on small particles.
Depending on the height in the planetary atmosphere where the scattering is most efficient, it can be due to scattering on H, H$_2$ or H$_2$O, or even on small dust grains.  If the latter is the case, the grain size should be $\le$0.5$\mu$m to have the Rayleigh type scattering in the blue. Such dust may be present in the atmospheres of hot Jupiters, e.g., in form of silicate granes \citep{rich07}. More accurate multi-color polarimetric measurements are needed to constrain the nature of scattering particles.

\begin{figure}
\resizebox{8.5cm}{!}{\includegraphics{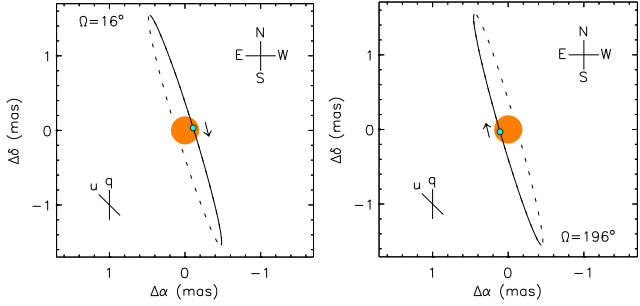}}
\caption{The orbit of HD189733b as projected on the sky. {\it Solid} and {\it dashed} lines indicate parts of the orbit in front of and behind the sky plane, respectively. The circle in the center depicts the star and the smaller circle on the orbit the planet. The direction of the orbital motion is indicated by the arrows. The reconstruction is made for $i$=180\degr$-$86\degr=94\degr\ and $\Omega$=16\degr\ (left) and 196\degr\ (right). If the inclination is 86\degr, the $\Omega$ values are to be swaped and the direction of the orbital revolution is to be made counterclockwise. Positive directions of Stokes $q$ and $u$ and orientations on the sky are also shown. [{\it See the electronic edition of the Journal for a color version of this figure.}]}
\label{fig:orb}
\end{figure}

Knowing all the orbital parameters of HD189733b allows us to depict its orbit as projected on the sky plane and indicate the direction of the orbital motion. This is shown in Fig.~\ref{fig:orb}. Such a plot is useful for future imaging or interferometric studies of the system as the coordinates of the planet can be predicted quite accurately. Since spectroscopy during a transit revealed that the sky projections of the stellar spin axis and the orbit normal are aligned to within a few degrees \citep{winn06}, the direction of the stellar axis becomes known as well.

\section{Conclusions}\label{sec:con}

The first direct detection of the visible light from an extrasolar planet allowed us to obtain fundamentally new information on the planet's orbit and scattering properties. We determined the orientation of the HD189733b orbit projected on the sky, which can be used for subsequent direct detections by, e.g., radio interferometry. Moreover, our independent estimates of the orbit inclination and eccentricity are in excellent agreement with the values determined previously from transits and radial velocities. In addition, we inferred that the planet has an extended atmosphere which efficiently scatters the stellar light in the blue. 

Our findings open the door to new opportunities for direct detections of extrasolar planets, both hot Jupiters and Earth-like, to a large degree independently of their mass and gravitational effect on the host star. Furthermore, until now probing exoplanetary atmospheres was largely limited to systems with transits, which are relatively rare events. Polarimetry provides us with a new prospect to detect directly the light from the planetary atmosphere outside transits. Thus, with polarimetric accuracy approaching the photon noise limit, direct studies of exoplanetary atmospheres in the visible at any orbital inclination become reality.

We are thankful to Jeff Kuhn for very useful discussions of the data analysis and results.
SVB acknowledges the EURYI (European Young Investigator) Award provided 
by the European Science Foundation (see www.esf.org/euryi) and SNF
grant PE002-104552.

\end{document}